\documentclass[twocolumn,showpacs,preprintnumbers,amsmath,amssymb]{revtex4}

\usepackage{graphicx}
\usepackage{dcolumn}
\usepackage{bm}
\usepackage{color}

\begin{document}

\preprint{APS/123-QED}

\title{Modulational instability, wave breaking and  formation of large scale dipoles
in the atmosphere}


\author{A. Iafrati$^1$}
\author{A. Babanin $^{2}$}
\author{M. Onorato$^{3,4}$}
\ 
\affiliation{
$^1$CNR-INSEAN - Italian Ship Model Basin - Roma, Italy;\\
$^2$ Swinburne Univ. Technology, Melbourne, Australia; \\
$^3$Dip. di Fisica , Universit\`{a} di
Torino, Via P. Giuria, 1 - Torino, 10125, Italy; \\
$^4$ INFN, Sezione di Torino, Via P. Giuria, 1 - Torino, 10125, Italy; 
}

\date{\today}

\begin{abstract}
In the present Letter we use the Direct Numerical Simulation (DNS) of the Navier-Stokes equation 
for a two-phase flow (water and air) to study the dynamics of the modulational 
instability of free surface waves
and its contribution to the interaction between ocean and
atmosphere.
If the steepness of the initial wave is large enough,  we observe 
a wave breaking and 
the formation of large scale dipole structures in the air. Because of the 
multiple steepening and breaking of the waves under unstable wave packets, 
a train of dipoles is released and propagate in the atmosphere at a height comparable 
with the wave length. 
The amount of energy dissipated by the breaker in water and air is considered 
and, contrary to expectations, we observe that the energy dissipation in air 
is larger than the one in the water. Possible consequences on the 
wave modelling and on the exchange of aerosols and gases between air and 
water are discussed.

\end{abstract}

\pacs{Valid PACS appear here}
\maketitle
The modulational instability, also known as the Benjamin Feir instability,
is  a well known universal phenomenon that takes place in many different fields of physics such 
as surface gravity waves, plasma physics, nonlinear optics (see the recent 
historical review
\cite{zakharov2009modulation}).
The basic idea consists in that a sufficiently steep sinusoidal wave may 
become unstable if perturbed by a long enough perturbation. It is a threshold 
mechanism, therefore, for example, for surface gravity waves in infinite 
water depth a wave is unstable if $2 \sqrt{2} k_0 A_0> \Delta k/k_0$, where 
$k_0$ is the wave number of the sinusoidal wave (carrier wave),  
$\Delta k$ is the wave number of the perturbation and $A_0$ is the amplitude
of the initial wave. 

The modulational instability has been discovered in the sixties and recently it has  received 
again attention because it has been recognized as 
a possible mechanism of formation of the rogue waves 
\cite{onorato01,janssen03}.
 The standard mathematical tool used to describe
such physical phenomena is the Nonlinear Schr\"ordinger equation (NLS), 
which is a weakly nonlinear, narrow band approximation of some primitive 
equation of motion. 
The beauty of such equation is that it is integrable and many analytical 
solutions can be written explicitly.  For example, breather solutions 
\cite{osborne00,dysthe99} have been considered
as prototypes of rogue waves; they have been observed in 
controlled experiments both in surface gravity waves and in nonlinear optics 
\cite{chabchoub2011rogue,clauss2011formation,kibler2010peregrine,kibler2012observation}.

Concerning the ocean waves, the studies on the modulational instability 
have been concentrated on the NLS dynamics and only more recently numerical 
computation of non-viscous, potential, fully nonlinear equations have been 
considered \cite{dyachenko2008formation,babanin2007predicting}.
However, so far none of the aforementioned literature has 
 ever considered the effect of the modulation instability on the fluid 
above the free surface. 
As far as we know this is the first attempt  in which the dynamics of air on 
water during the modulation process is investigated. 
This has been possible by simulating the Navier-Stokes equation for a two 
phase flow. This approach allows us to investigate conditions which are 
beyond the formal applicability of the NLS equation: for example, it is well 
known that if the initial wave steepness is large enough, the NLS equation 
is not able to describe the dynamics because the breaking of the wave takes 
place \cite{henderson1999unsteady,babanin2011wave}.
For steep waves and particularly those close to the breaking onset, vorticity is generated 
by viscous effects and by the topological change of the interface in
case of bubble entrainment processes (this dynamics cannot be 
described by potential models).
Breaking of surface waves, as an oceanic phenomenon   \cite{babanin2011breaking}, is 
important across a very broad range of applications related to 
wave dynamics, atmospheric boundary layer,
air-sea-interactions, upper ocean turbulence mixing, with respective
connections to the large-scale processes including ocean circulation,
weather and climate \cite{cavaleri2011wind}. 
Modulational instability and breaking has become also relevant in engineering applications 
\cite{Clauss2012Application}
such as naval architecture, structural design of offshore developments, 
marine transportation, navigation, among many others.

In the present work, the two-fluid flow of air and water is approximated 
as that of a  single incompressible fluid with density and viscosity
smoothly varying across the interface. 
The continuity and momentum equations (Navier-Stokes) are solved 
in generalized coordinates \cite{iafrati2009numerical, iafrati2011energy}.
The variation of the fluid properties and the surface tension forces is 
spread across a small neighborhood of the interface. 
The interface between air and water is captured as the zero 
level-set of a signed  distance from the interface
$d({\boldsymbol x},t) $ which, at $t=0$, is initialized by assuming $d
> 0$ in water, $d < 0$ in air.
Physical fluid properties are assumed to be related to $d$ by the equation:
\begin{equation} \label{15}
f(d) = f_a + (f_w-f_a) H_{\delta}(d)
\end{equation}
where $H_{\delta}(d)$ is a smooth step function and the parameter $\delta$ is chosen so that the
density and viscosity jumps are spread across some grid cells
\cite{iafrati2005free}.
The distance function is advected in time with the flow as a non-diffusive 
scalar by using the equation
\begin{equation} \label{17}
\frac{\partial d}{\partial t} + {\boldsymbol u} \cdot \nabla d = 0,
\end{equation}
and the interface is located as the $d=0$ level. Hence,  the distance
is reinitialized.

In our simulations we consider the standard modulational instability process as the one 
produced for example in the experimental work in \cite{tulin99}. 
The initial surface elevation  is characterized by a perturbed 
sinusoidal free surface elevation of the form:
\begin{equation}
\label{ele0}
\eta(x,t=0) = A_0 \cos(k_0 x) + A_1( \cos(k^+ x) + \cos(k^- x)),
\end{equation}
where $k_0$ is the wave number of the carrier wave, $k^\pm =
k_0 \pm \Delta k$ with $\Delta k$ the wavenumber of the perturbation.
The simulations presented hereafter are characterized by a $\epsilon_0=k_0 A_0$ 
that is varied
from $0.1$ to $0.18$, with a  step $0.02$.
 The sideband components are placed at $\Delta k = k_0/5$ and their amplitude 
is $A_1 = 0.1 A_0$.  It is worth noticing that the conditions are essentially 
similar to those used in \cite{babanin2007predicting,song2002}  and corresponds to 
the early stages of an Akhmediev breather \cite{akhmediev1987exact}. 

Because of the typical time scale of the modulational
instability is of the order of 100 periods, the Navier-Stokes simulation 
is expensive
and for the initial development of the instability a standard potential code 
is used.
The Navier-Stokes simulation is then initialized with solution from the potential flow.
For convenience, results are presented in dimensional form.
Simulations are carried out for carrier wave of  wavelength  $\lambda_0
= 0.60 $ m, with $g = 9.81$ m s$^{-1}$. 
The computational domain which spans horizontally from $x$ = -1.5 m to $x$ = 1.5 m and 
vertically from $y $= -2 m up to 0.6 m above the still water level. The domain is uniformly discretized in the horizontal direction with $\Delta x$ = 1/1024 m. Vertically, the grid spacing is uniform, and equal to 
$\Delta x$, from $y$ = -0.15 to 0.15 m whereas it grows geometrically by a factor $\alpha$= 1.03 towards the upper and lower boundaries. This gives a total of  3072 $\times$ 672 grid cells. 
The total thickness of the transition region is 0.01 m, so that the density jump is spread across about 10 grid cells.
Although neglected in the potential 
flow model, surface tension effects are considered in the Navier-Stokes
solution for which the surface tension coefficient is assumed to be that
in standard conditions $\sigma = 0.073$ N m$^{-1}$.
In the two-fluid modelling, the densities of air and water are the standard 
ones, $\varrho_w = 1000 $ kg m$^{-3}$ and $\varrho_a = 1.25 $ kg m$^{-3}$. 
The values of the dynamic viscosities in  water
and air are $\mu_w = 10^{-3} $ kg m$^{-1}$ s$^{-1}$ and 
$\mu_a = 1.8 \, 10^{-5}$ kg m$^{-1}$ s$^{-1}$, respectively.

\begin{figure}
\centerline{\includegraphics[width=8cm]{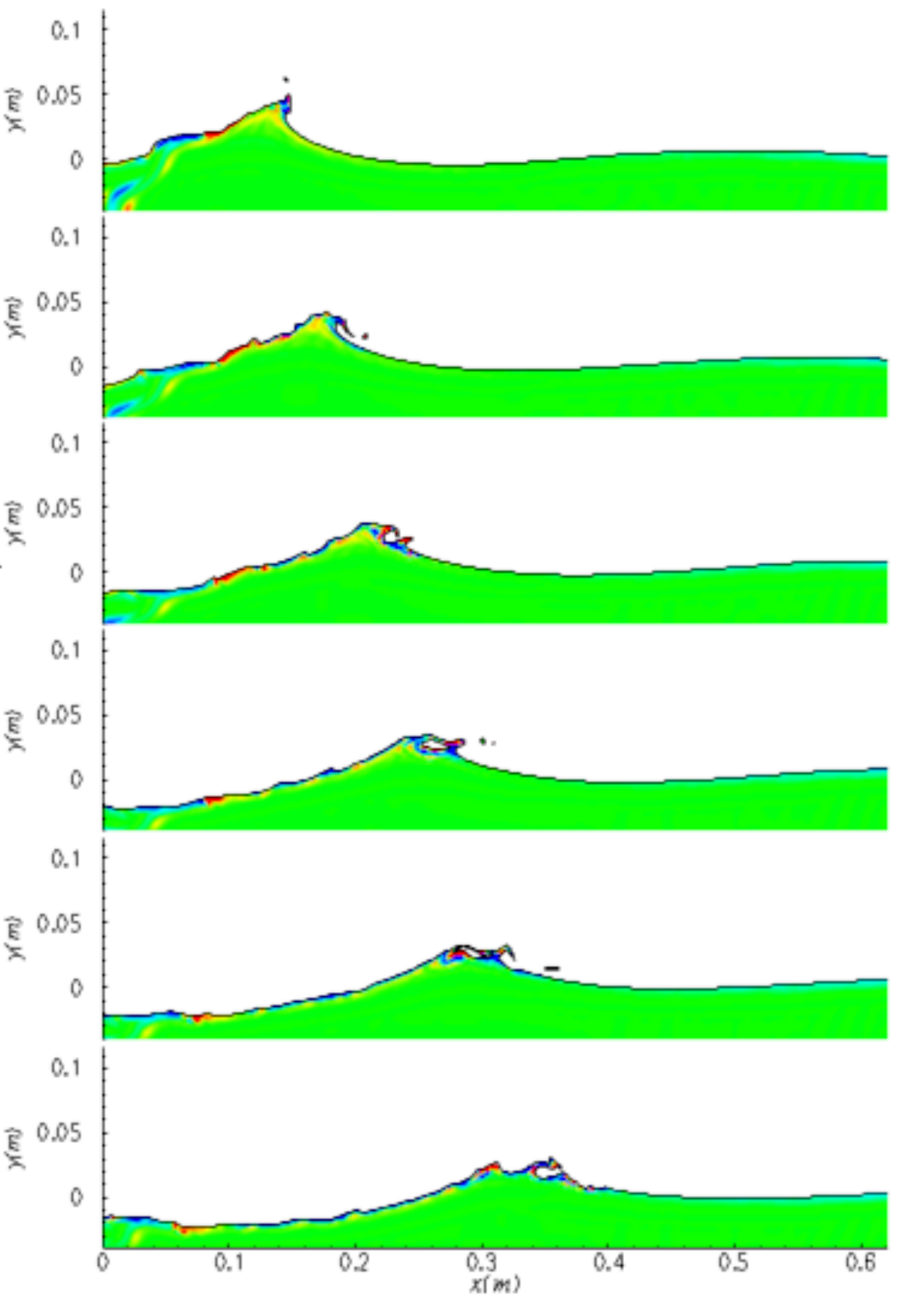}}
 \caption{Breaking event for initial steepness $\epsilon_0$=  0.18. 
The color corresponds to the vorticity field: green corresponds to zero 
vorticity, blue to -20  $sec^{-1}$ and red to 20 $sec^{-1}$. }
 \label{breaking_1}
 \end{figure}

The process observed in the simulation corresponds to the standard modulational instability (exponential growth of the side bands) up to 
the point where the wave group reaches its strongly nonlinear regime and eventually wave breaking is observed.
We first concentrate our attention to the breaking event:
in figure \ref{breaking_1}  an example for steepness $\epsilon_0=k_0A_0$= 0.18 
of the simulation of the 
wave breaking resulting from the modulational instability is shown; 
the formation of the first jet is observed which entraps the air. The jet then 
bounces on the free surface creating a second air bubble. Some droplets 
of water in air are also visible; a small amount of vorticity is also released 
beneath the surface. 
 
\begin{figure}
\centerline{\includegraphics[width=8cm]{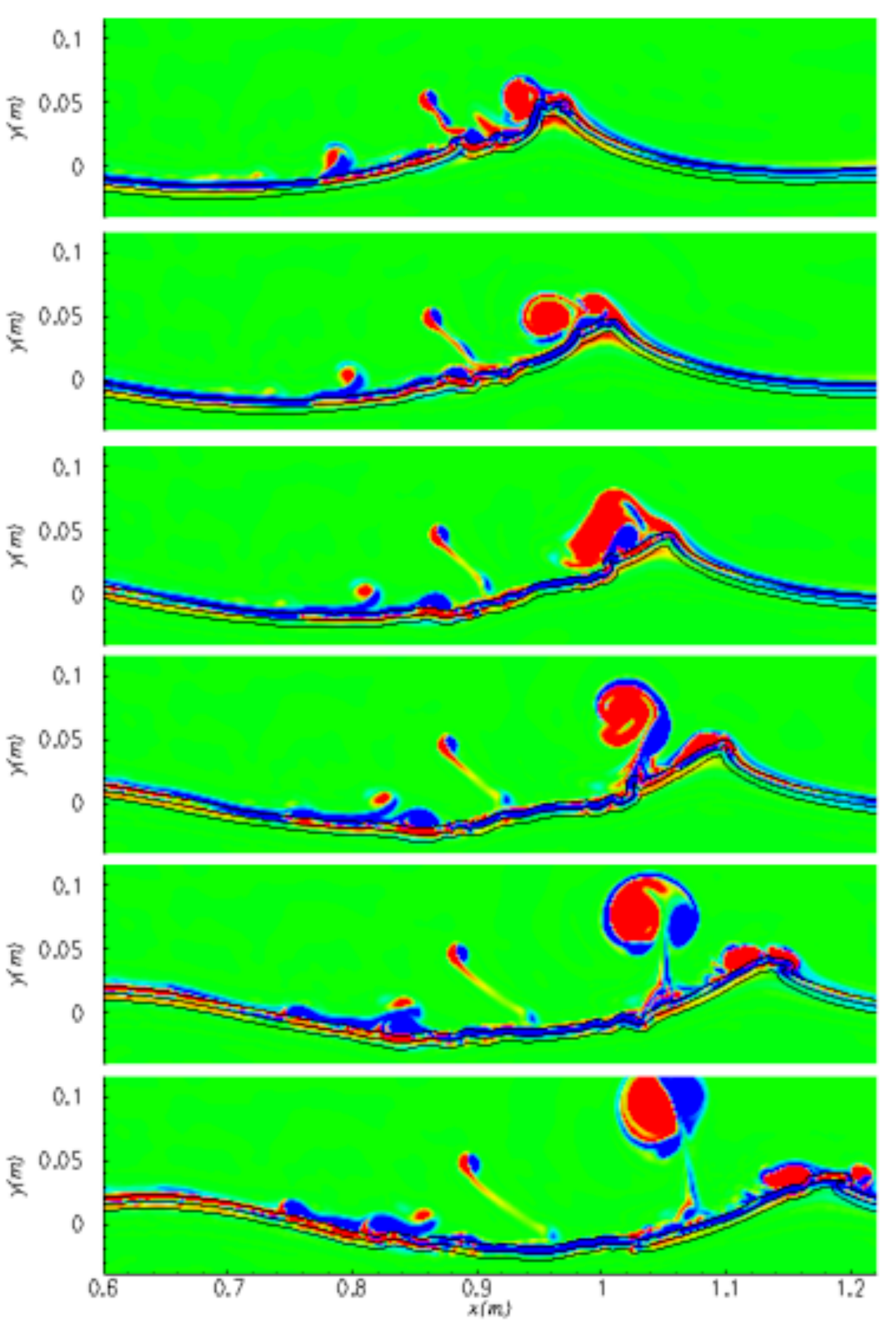}}
 \caption{Sequences of formation of a dipole in air as a consequence of 
the wave breaking. Results refer to an initial steepness 
$\epsilon_0$= 0.16}
 \label{roll_up}
 \end{figure}

It is of particular interest to  focus the attention on the 
dynamics of the air flow. In figure \ref{roll_up}
we show a sequence  of snap-shots of the water and air domain where the 
formation and detachment of a dipole structure 
is observed (the simulation is performed for initial steepness of $\epsilon_0=0.16$):
the rather fast steepening of the wave profile and the wave breaking causes the air flow to
separate from the crest giving a rise to large, positive, vorticity 
structure (positive vorticity is in red). The interaction of this vortex 
structure with the free
surface leads to the formation of a secondary vorticity structure of opposite 
sign, which eventually detaches from the free surface and forms a dipole 
which moves
under the self-induced velocity (such phenomenon 
is not observed if the wave breaking process does not take place: we have 
performed a numerical simulation with 
$\epsilon_0=0.1$ which does not lead to a breaking event and no evident 
 dipole structures have been observed).

\begin{figure}
\centerline{\includegraphics[width=9
cm]{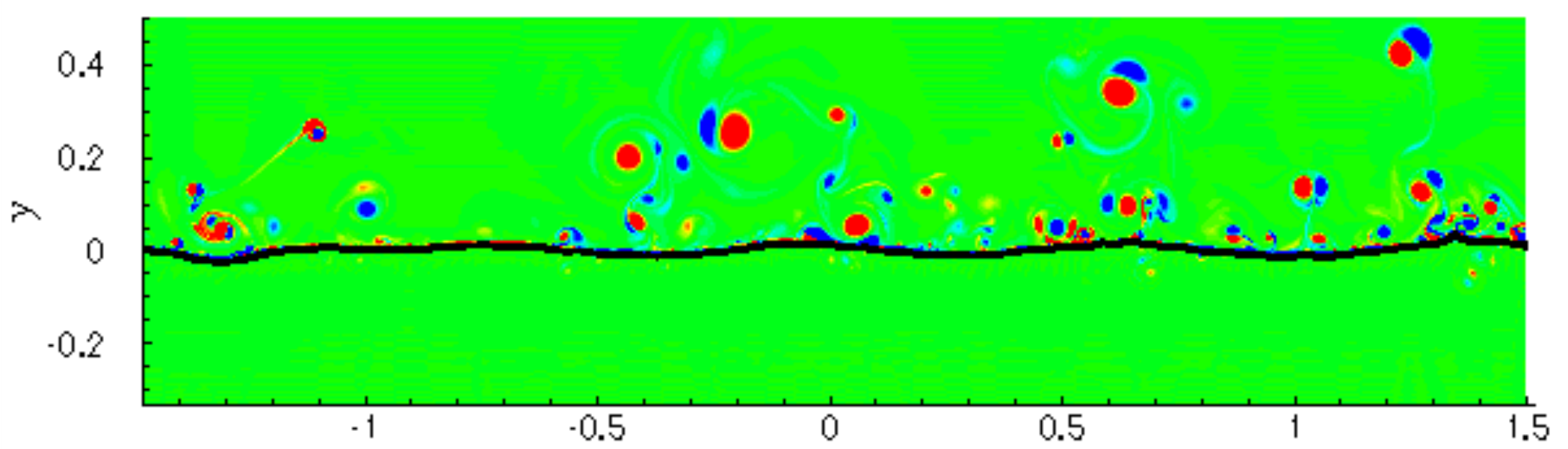}}
 \caption{Vorticity field in a portion of the computational domain for initial steepness $\epsilon_0=0.18$. The scale of vorticity is as described in the label of figure \ref{breaking_1}
  The supplemental material contains the animation of the simulation (file: animation18.avi). 
}
 \label{dipoles}
 \end{figure}

Because the group velocity is half the phase velocity, each single wave 
that passes 
below the group (at its maximum height), breaks. The result is that 
a series of dipoles are released into the atmosphere as shown in figure 
\ref{dipoles}.  Vortices 
of various sizes and dipoles are clearly observable in the domain.
Two things should be noted: i)  the height of the highest dipoles is of the order of 
the wavelength; ii) large amount of vorticity 
 is observed in the air and not in the water.
 \begin{figure}
\centerline{\includegraphics[width=9cm]{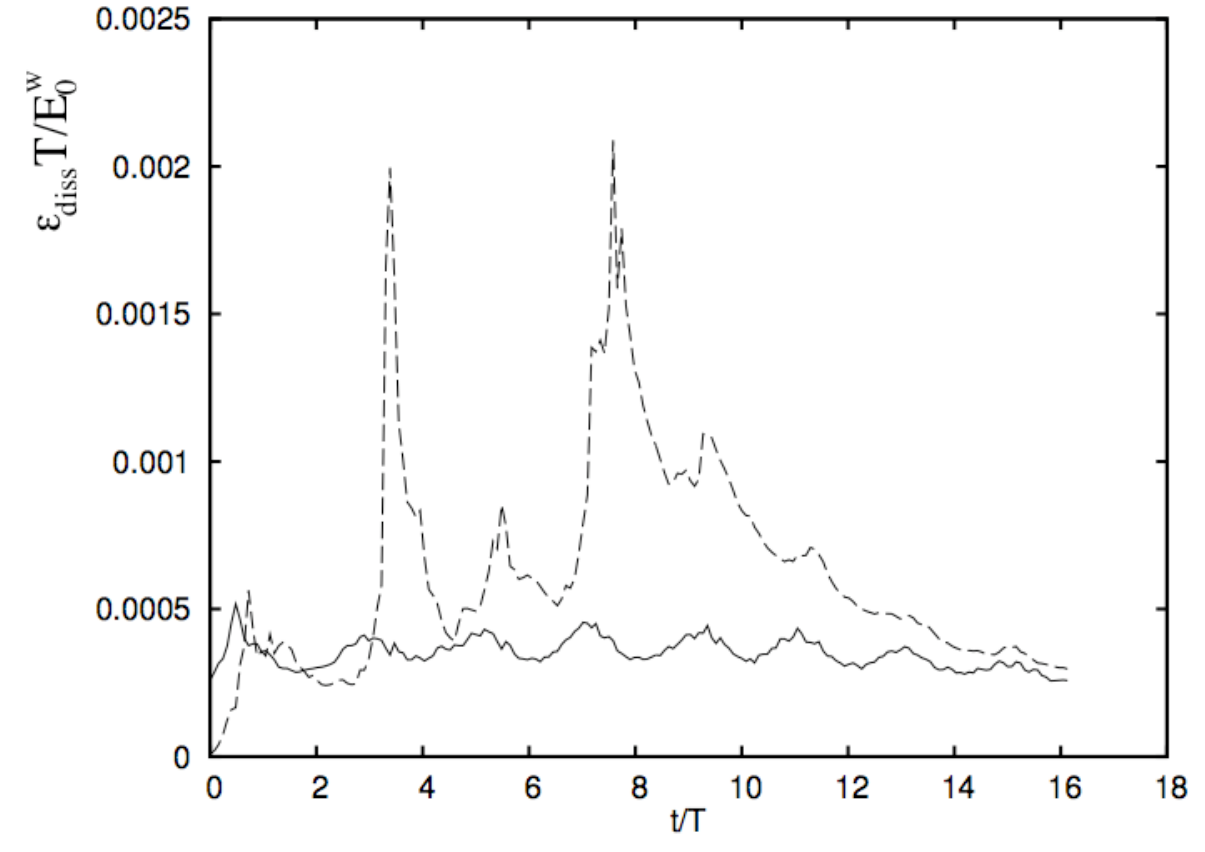}}
 \caption{Dissipation nondimensionalized with period $T$ and 
 total energy in the water, $E_0^w$ a $t=0$ as a function of nondimensional time for water (solid line) and air (dashed line). The initial steepness is 0.12}
 \label{dissipation1}
 \end{figure}
 \begin{figure}
\centerline{\includegraphics[width=9cm]{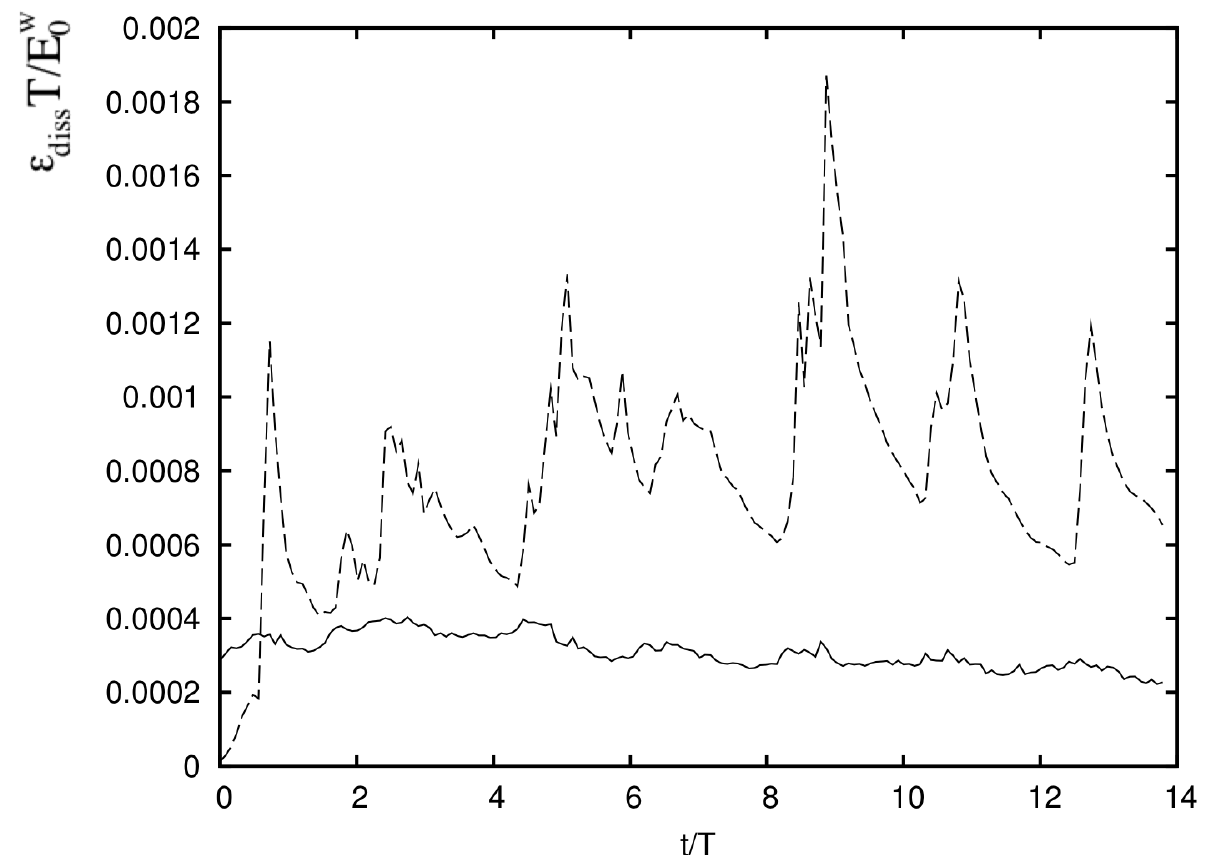}}
 \caption{Dissipation nondimensionalized with period $T$ and 
 total energy in the water, $E_0^w$ a $t=0$ as a function of nondimensional time for water (solid line) and air (dashed line). The initial steepness is 0.18}
 \label{dissipation2}
 \end{figure} 
One major question to be answered, especially in the 
spirit of modelling the dissipation term in the wave forecasting models
\cite{komen94,cavaleri2007wave} is the amount of energy dissipated  during 
a wave breaking or a sequence of breaking events. Therefore, a
quantitative estimate of the dissipated energy both in air and water 
 can be obtained by integrating the viscous
stresses over the air and water domain:
\begin{equation}
\label{eqdnw}
\epsilon_{diss}^{w}(t) = \int_{d \ge \delta} 2 \mu e_{ij} \frac{\partial u_i}
{\partial x_j} \mbox{
d}x \mbox{ d}y, 
\end{equation}
\begin{equation}
\label{eqdna}
\epsilon_{diss}^{a}(t) = \int_{d < -\delta} 2 \mu e_{ij} \frac{\partial u_i}
{\partial x_j} \mbox{
d}x \mbox{ d}y, 
\end{equation}
where $e_{ij}$ is the symmetric part of the strain tensor.
In figure \ref{dissipation1} and \ref{dissipation2} we show the dissipation 
function normalized by the initial energy of the water and wave period, $T$,
as a function of time, nodimensionalized by $T$, for a simulation 
with steepness $\epsilon_0=0.12$ and $\epsilon_0=0.18$, respectivelly; the origin of the time axis is set to the
time at which the Navier Stokes simulations takes over the simulation with 
the potential code. 
The occurrence of spikes in the energy content in the air indicates that an
energy fraction is transferred to the air and is successively dissipated by
the viscous stresses. The multiple peaks correspond to the 
the multiple breaking that occurs during the modulational
instability process (as mentioned before, the group is slower than the 
phase velocity and each wave breaks as it passes below it). 
Similar plots are observed for larger steepness.

\begin{figure}
\centerline{\includegraphics[width=8cm]{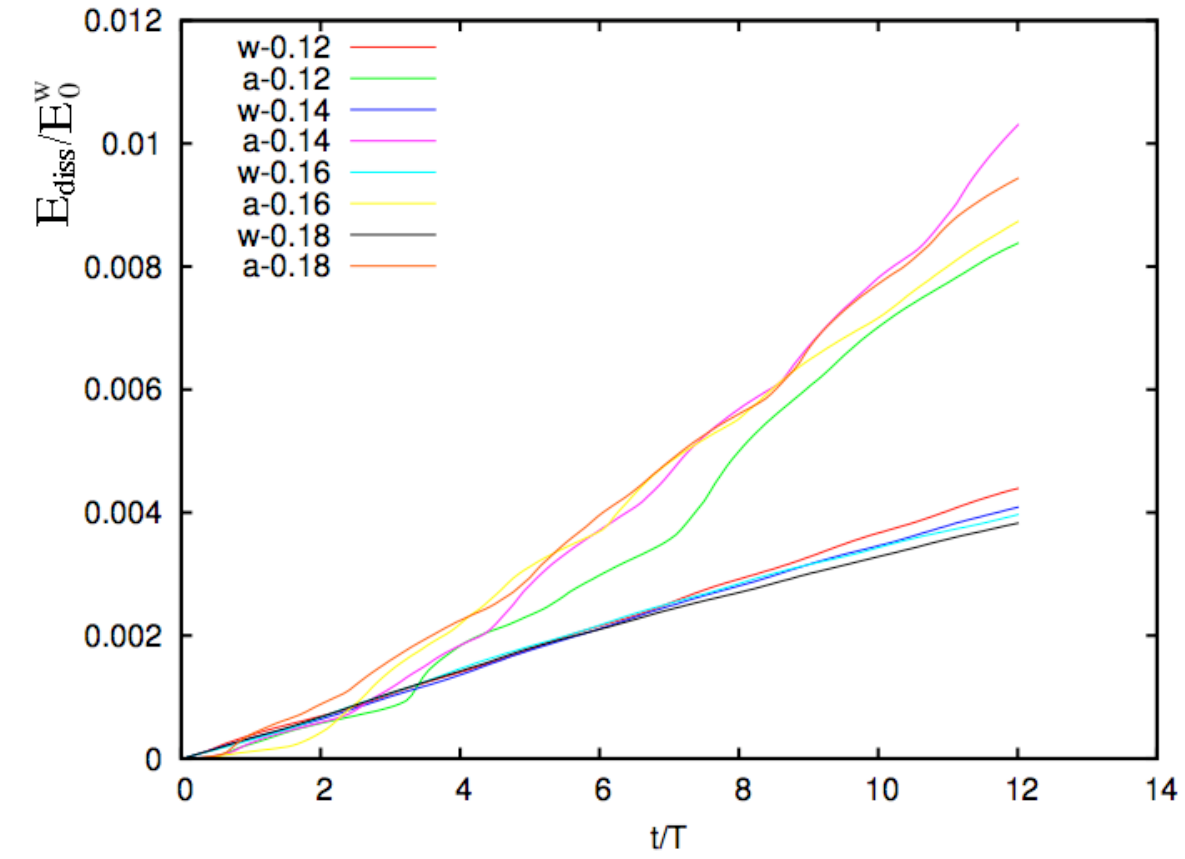}}
 \caption{Integrated dissipation for different values of the 
 steepness 
 in water (w) and air (a).}
 \label{int_disspation}
 \end{figure}

In order to show how relevant is the energy dissipation in the air 
in comparison to the corresponding one in water, we
 consider the following integrated quantity:
\begin{equation}
E_{diss}(t)=\int_0^t \epsilon_{diss}(t') dt'.
\end{equation}
The integral is considered for both for water and air.
Histories of the integrals in time of the viscous dissipation terms 
in the two 
media are shown  in figure \ref{int_disspation}. 
It is rather interesting to see that, in nondimensional
form, the solutions for the four different cases  almost overlap, 
with a total dissipation in the air about three times that in
water.

{\it Discussion -} Modulated waves of different initial steepnesses have 
been analyzed using the two phase flow NS equation. 
For initial steepness larger or equal to 0.12,  multiple breaking events have been observed.
Spray is the natural consequence of the wave breaking. Droplets of 
water  are thrown in the air; some of these particles are so small (aerosols) 
that they can remain in the air for a very long time, forming condensation 
nuclei for clouds and affecting incoming solar radiation and therefore relevant for 
climate modelling.
Vortices, observed in our simulations, can in principle transport 
aerosols (not resolved in our simulations) up to the height of the wave 
lengths (this can be even underestimated because of the presence of the solid boundary at the top of the computational domain).  
Quantitative analyses of the energy contents and of the viscous dissipation
term have been provided. Even bearing in mind the limits of the
numerical scheme, the results indicate that, due to the highly rotational 
flow in air, the dissipation of the energy is mostly concentrated in the
air side. We stress that it is a common practice to estimate the energy loss
due to a wave breaking by looking at the amount of energy dissipate in the water,
see for example \cite{terray1996estimates}. Such measurement are the bases for the construction
of the dissipation function in the operational wave forecasting modelling \cite{cavaleri2007wave}.
 If the modulational instability 
is responsible of the wave breaking in the ocean, then such results would
clearly lead to an underestimation of the the total energy dissipated.
The present work represents the first attempt to study the modulational instability
starting from the Navier-Stokes equation and its effects on water and air.
Clearly, it has a number of limitation due to the 
heaviness of the computation. Probably the most important one is that we have 
assumed that waves are long crested and 
the fluid domain is 2-dimensional. In reality, we expect that 3D effects can take place and dipoles
and vortices can become unstable. 
We also underline that our simulations correspond to the propagation of
 waves without the presence of external wind:  what would be the consequences of a turbulent wind on the generation of 
 vorticity during breaking event is under investigation. 

 {\bf Acknowledgments} 
M.O. has been funded by EU, project EXTREME SEAS (SCP8-GA-2009-234175) and by ONR grant N000141010991.
Dr. Proment and Dr. Giulinico are acknowledged for discussion.
The work by A.I. has been done within the RITMARE research project.
\bibliography{references}
\end{document}